# Characterization of the Reflectivity of Various Black Materials

Jennifer L. Marshall*, Patrick Williams, Jean-Philippe Rheault, Travis Prochaska, Richard D. Allen, D. L. DePoy

Department of Physics and Astronomy
Texas A&M University, 4242 TAMU, College Station, TX 77843-4242 USA

## ABSTRACT

We present total and specular reflectance measurements of various materials that are commonly (and uncommonly) used to provide baffling and/or to minimize the effect of stray light in optical systems. More specifically, we investigate the advantage of using certain black surfaces and their role in suppressing stray light on detectors in optical systems. We measure the total reflectance of the samples over a broad wavelength range ($250 < \lambda < 2500$ nm) that is of interest to astronomical instruments in the ultraviolet, visible, and near-infrared regimes. Additionally, we use a helium-neon laser to measure the specular reflectance of the samples at various angles. Finally, we compare these two measurements and derive the specular fraction for each sample.

*marshall@physics.tamu.edu; phone 1 979 862-2782; fax 1 979 845-2768; http://instrumentation.tamu.edu

## 1. INTRODUCTION

An important consideration in the design and construction of optical and infrared astronomical instruments is the minimization of stray and scattered light within the instrument. Historically, a wide range of materials have been used to minimize unwanted reflections within an instrument, from simple household materials to specialized treatments such as black anodization of aluminum. In recent years, a wide variety of modern sophisticated materials, paints, and coatings have been developed specifically for this purpose. Nearly all of these materials appear "black" to the human eye; this property, however, does not guarantee that the materials function adequately as blackout materials at the wide range of wavelengths studied by astronomical instruments. In this paper we present measurements of the total and specular reflectance of various black materials that may be used to minimize unwanted reflections over a wavelength range relevant to optical and infrared astronomical instrumentation ($250 < \lambda < 2500$ nm).

## 2. MATERIALS TESTED

We have tested a range of materials that may be used to baffle the optical path and minimize stray light in modern astronomical instruments. Specifically, we tested metal (aluminum and steel) samples treated with a range of techniques, from anodization to simple black paints. We also analyzed the reflectivity of blackout materials that might be used in an effort to mitigate stray light in an instrument. These range from high end commercial products specifically designed to reflect almost no light over a wide wavelength range, to household materials (e.g. permanent marker and black construction paper) that in our experience have been used for similar purpose in existing astronomical instruments.

We prepared samples of both metal and non-metal materials. Each sample was sized to fit in the measurement instrument, which accepts a 2 inch square sample that is no more than 0.75 inch thick. The complete list of samples tested is provided in Table 1.

### 2.1 Metal samples

We prepared samples of 6061 aluminum, cast aluminum, stainless steel, and two types of Invar. "Thick" Invar refers to regular cast Invar and "thin" Invar refers to cold rolled sheet Invar.

Table 1. Samples tested.

| **Sample** | **Description** |
|---|---|
| A.R.B | Raw 6061 Aluminum, Anodized (MIL-A-8625, Type II, Class 2, Black) |
| A.M.B | Machined 6061 Aluminum, Anodized (MIL-A-8625, Type II, Class 2, Black) |
| A.P.B | Polished 6061 Aluminum, Anodized (MIL-A-8625, Type II, Class 2, Black) |
| A.B.B | Bead-Blasted 6061 Aluminum, Anodized (MIL-A-8625, Type II, Class 2, Black) |
| A.R.H | Raw 6061 Aluminum, Anodized (MIL-A-8625, Type III, Class 1, Non-dyed) |
| A.M.H | Machined 6061 Aluminum, Anodized (MIL-A-8625, Type III, Class 1, Non-dyed) |
| A.P.H | Polished 6061 Aluminum, Anodized (MIL-A-8625, Type III, Class 1, Non-dyed) |
| A.B.H | Bead-Blasted 6061 Aluminum, Anodized (MIL-A-8625, Type III, Class 1, Non-dyed) |
| C.R.B | Raw Cast Aluminum, Anodized (MIL-A-8625, Type II, Class 2, Black) |
| C.M.B | Machined Cast Aluminum, Anodized (MIL-A-8625, Type II, Class 2, Black) |
| C.P.B | PolishedCast Aluminum, Anodized (MIL-A-8625, Type II, Class 2, Black) |
| C.B.B | Bead-Blasted Cast Aluminum, Anodized (MIL-A-8625, Type II, Class 2, Black) |
| C.R.H | Raw Cast Aluminum, Anodized (MIL-A-8625, Type III, Class 1, Non-dyed) |
| C.M.H | Machined Cast Aluminum, Anodized (MIL-A-8625, Type III, Class 1, Non-dyed) |
| C.P.H | Polished Cast Aluminum, Anodized (MIL-A-8625, Type III, Class 1, Non-dyed) |
| C.B.H | Bead-Blasted Cast Aluminum, Anodized (MIL-A-8625, Type III, Class 1, Non-dyed) |
| I.R.N | Raw Invar (thick), Electroless Nickel Coat (MIL-C-26074) |
| I.M.N | Machined Invar (thick), Electroless Nickel Coat (MIL-C-26074) |
| I.P.N | Polished Invar (thick), Electroless Nickel Coat (MIL-C-26074) |
| I.R.N | Raw Invar (thin), Electroless Nickel Coat (MIL-C-26074) |
| I.P.N | Polished Invar (thin), Electroless Nickel Coat (MIL-C-26074) |
| S.R.N | Raw Stainless Steel, Electroless Nickel Coat (MIL-C-26074) |
| S.M.N | Machined Stainless Steel, Electroless Nickel Coat (MIL-C-26074) |
| S.P.N | Polished Stainless Steel, Electroless Nickel Coat (MIL-C-26074) |
| Permanent Marker on Aluminum | 6061 Aluminum coated with Sharpie black permanent marker |
| Weathered Permanent Marker on Aluminum | 6061 Aluminum coated with Sharpie black permanent marker left outside from 6 August 2012 to 27 August 2012 |
| Black Masking Tape | Thor Labs T743-2.0 |
| Black Aluminum Foil | Thor Labs BKF12 |
| Blackout Fabric | Thor Labs BK5 |
| Black Foam Board | Black foam board purchased at an arts and craft store. |
| Black Permanent Marker | Cardboard sample coated in black permanent marker (Sharpie). |
| Black Dry Erase | Cardboard sample coated in black dry erase marker (Expo). |
| Black Duct Tape | Cardboard sample covered in black duct tape. |
| Black Velcro | Cardboard sample covered in adhesive velcro (fuzzy side). |
| Black Electrical Tape | Cardboard sample covered in standard black electrical tape. |
| Black Felt | Black felt sample purchased at an arts and craft store. |
| Black Construction Paper | Typical black construction paper. |
| Valspar | Spray can Valspar premium enamel, flat and fast drying, for interior/exterior and ideal for wood, metal, and more. |
| Rustoleum | Spray can Rust-oleum specialty high heat tough protective enamel for grills, wood stoves. |
| Spraypaint | Generic spray paint for interior/exterior and fast drying. |
| Acktar Fractal Black | Black oxide optical coating sample from Acktar Advanced Coatings. |
| Acktar Metal Velvet | Black oxide optical coating sample from Acktar Advanced Coatings. |
| Rhodes LP Polisher | JH Rhodes polishing material. |
| Flock 55 | Edmund Optics Black Out Material Flock 55 |
| Flock 65 | Edmund Optics Black Out Material Flock 65 |

Before coating, metal surfaces were either left untreated (“raw”), or had their surfaces prepared by machining a flat plane on the surface of the sample, bead-blasting the surface of the sample, or were “polished” by sanding the faces of the samples in increasing grit sizes of 500, 1000, 1500, and 2000.

Metal samples were treated with a variety of coatings: aluminum samples were treated with black anodization (MIL-A-8625, Type II, Class 2, Black) and hardcoat anodization (MIL-A-8625, Type III, Class 1, Non-dyed), while steel samples were treated with electroless nickel plating (MIL-C-26074).

In Table 1 the samples are identified using a 3-letter ID system separated by periods. The first letter identifies the metal type (C: cast aluminum, A: 6061 aluminum, I: Invar, S: stainless steel). The second letter signifies the initial metal treatment (R: raw, P: Polished, M: machined, B: Bead-Blasted). Finally, the last letter identifies the coating treatment of the sample (B: Black-Dye Anodization, H: Hardcoat Non-Dyed Anodization, N: Electroless Nickel Coating). For example, a sample identified as C.R.B. is Cast Aluminum, Raw, with a Black-Dye Anodization.

### 2.2 Non-metal samples

We also prepared samples of materials that could be used in astronomical instruments to baffle light or otherwise minimize stray or scattered light. These materials range from household or art supply items to specially designed high-end blackout materials and coatings. The list of the non-metal materials tested is also provided in Table 1.

## 3. TOTAL REFLECTANCE MEASUREMENTS

### 3.1 Instrumental setup

We used the Hitachi High-Tech U-4100 UV-Visible-NIR Spectrophotometer in the Materials Characterization Facility (MCF) at Texas A&M University in order to obtain reflectance profiles for the samples.

The U-4100 dual beam spectrophotometer uses two different lamps to cover a wide range of wavelengths. For the far-UV ($\lambda<345$ nm), the U-4100 uses a deuterium lamp; the system uses a tungsten lamp for UV, visible, and near-IR measurements. The layout of the U-4100 includes monochromators, beam splitters, mirrors, focusing lenses, and detectors which can be used to analyze liquid or solid samples. With this system we measured precise reflectance values at each wavelength (in 1 nm steps) for the wavelength range $250 < \lambda < 2500$ nm.

Figure 1 shows the instrumental setup of the Hitachi High-Tech U-4100 UV-Visible-NIR Spectrophotometer used at the MCF at Texas A&M. The reference and test sample are placed in the 6 o'clock and 3 o'clock positions of the integrating sphere, respectively. The data acquisition procedure involves obtaining a baseline measurement at each wavelength of the reference BaSO4 wafers (~100% reflectance) in both the reference and sample slots of the dual beam spectrophotometer. We then measure a second reference sample having 5% reflectivity (Labsphere SRS-05), and measure the reflectivity of the test sample. We compare the 5% reflectance reference sample to the values provided by the manufacturer and use this ratio to construct the absolute reflectivity of the test sample as a function of wavelength.

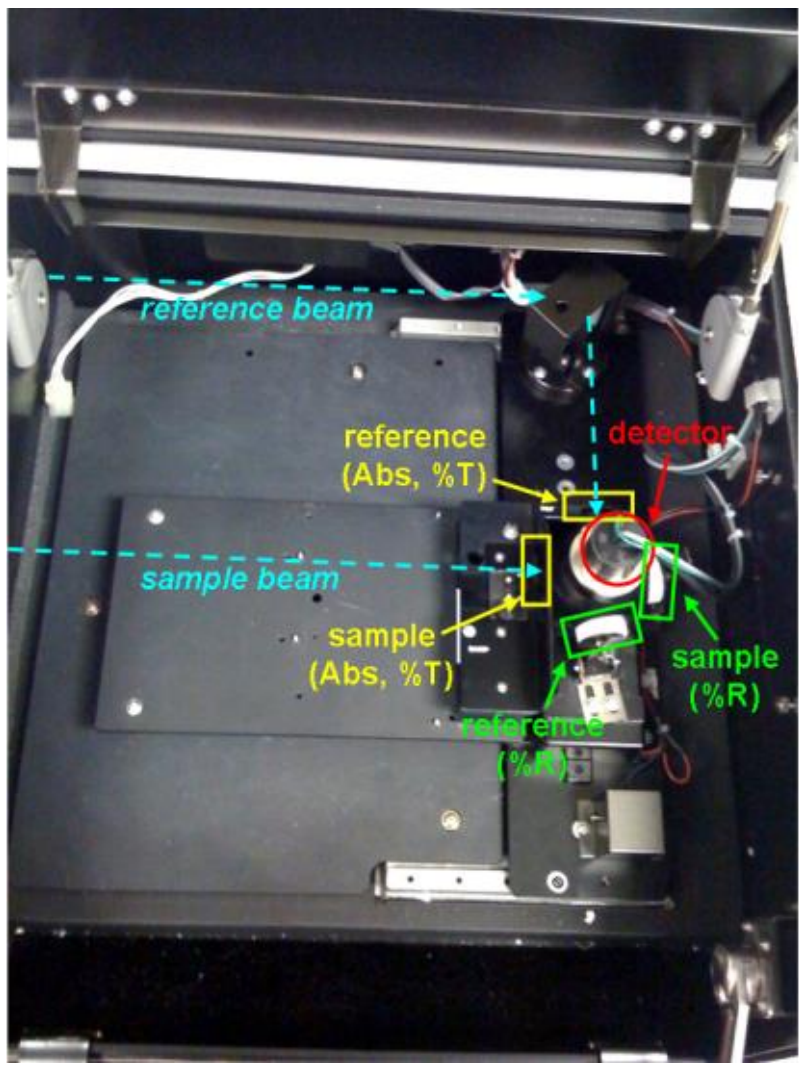

Figure 1. Instrumental setup of the Hitachi U-4100 spectrophotometer.

## 3.2 Total reflectance measurements

In the figures below we present the measured total (diffuse and specular) reflectance of each of the materials studied.

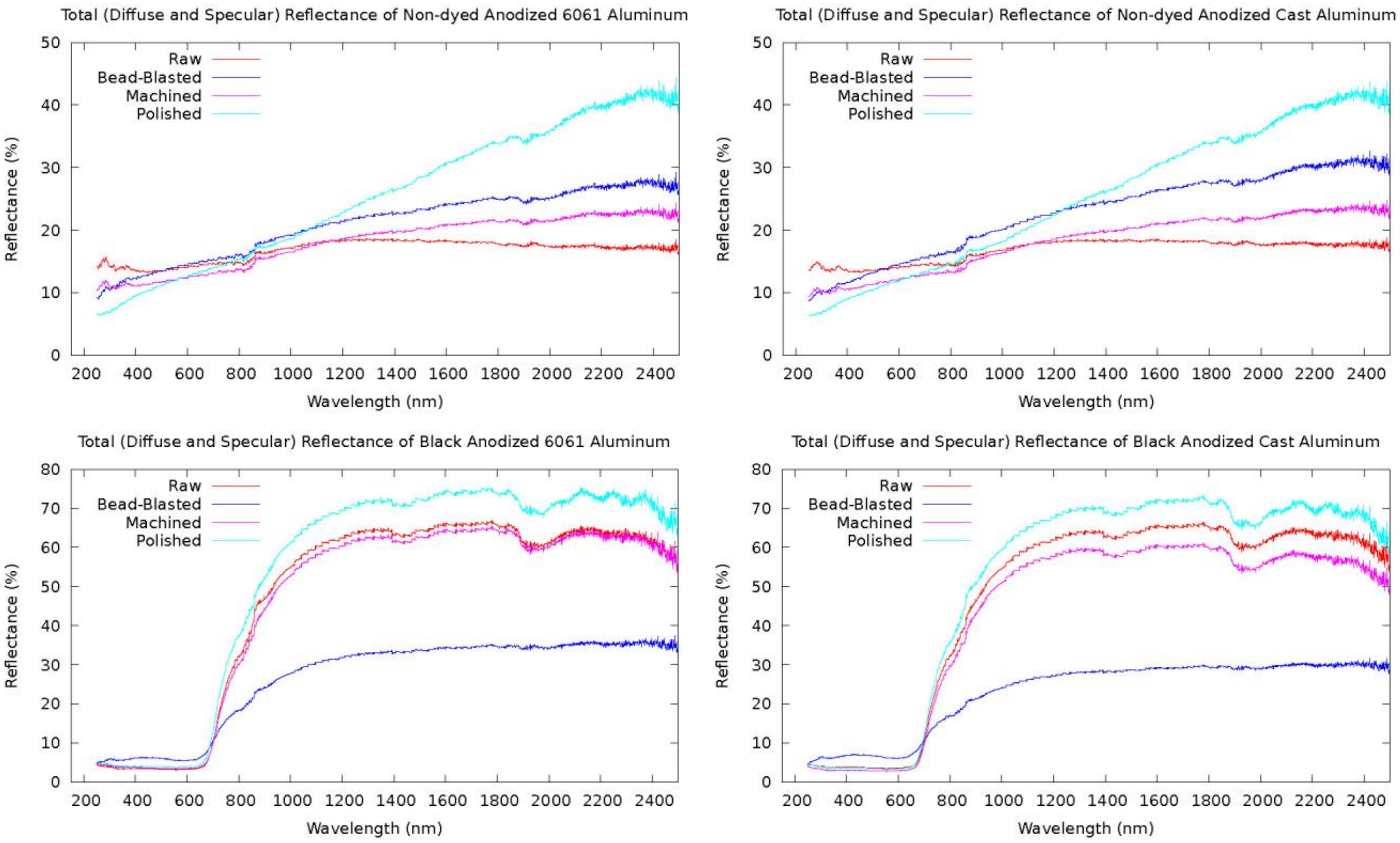

Figure 2. Total reflectance of aluminum samples.

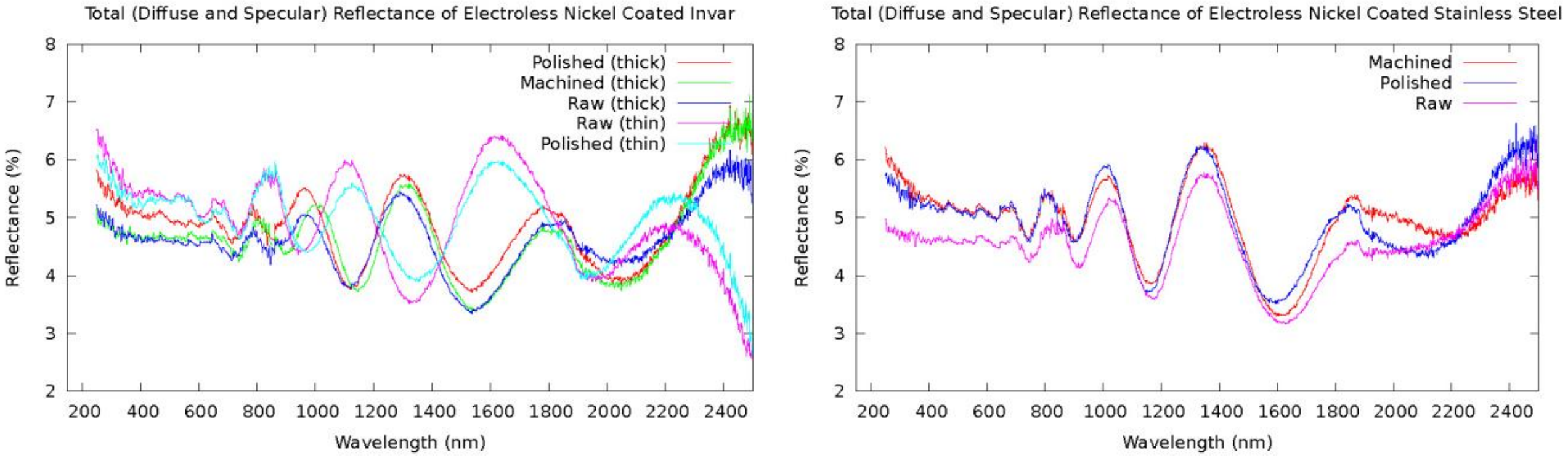

Figure 3. Total reflectance of Invar and stainless steel samples.

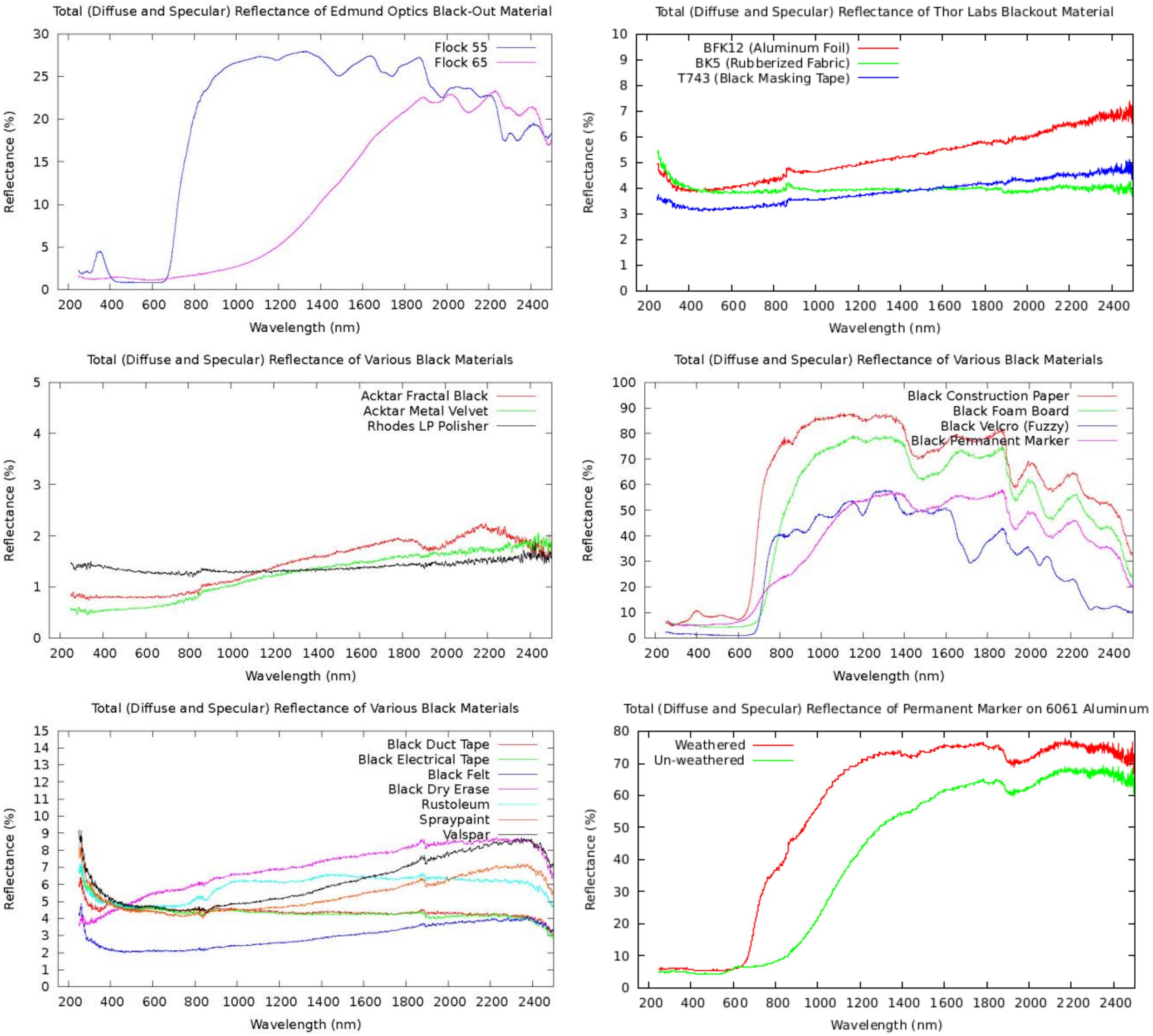

Figure 4. Total reflectance of various commercial and household blackout and baffling materials.

## 4. SPECULAR REFLECTANCE

We also measured the specular reflectance for many of the samples.

### 4.1 Instrumental setup

In this measurement we use a Helium-Neon laser reflected by the surface of each of our samples at 10°, 22°, and 44°. We use a Gentec Photo-Detector (PH100-SiUV; S/N: 181951) to measure the specular intensity at distance of approximately 1 meter from the sample.

Figure 5 shows the specular reflectance setup constructed on an optics bench in the Munnerlyn Astronomical Instrumentation Laboratory at Texas A&M University. All measurements were taken in a dark room environment.

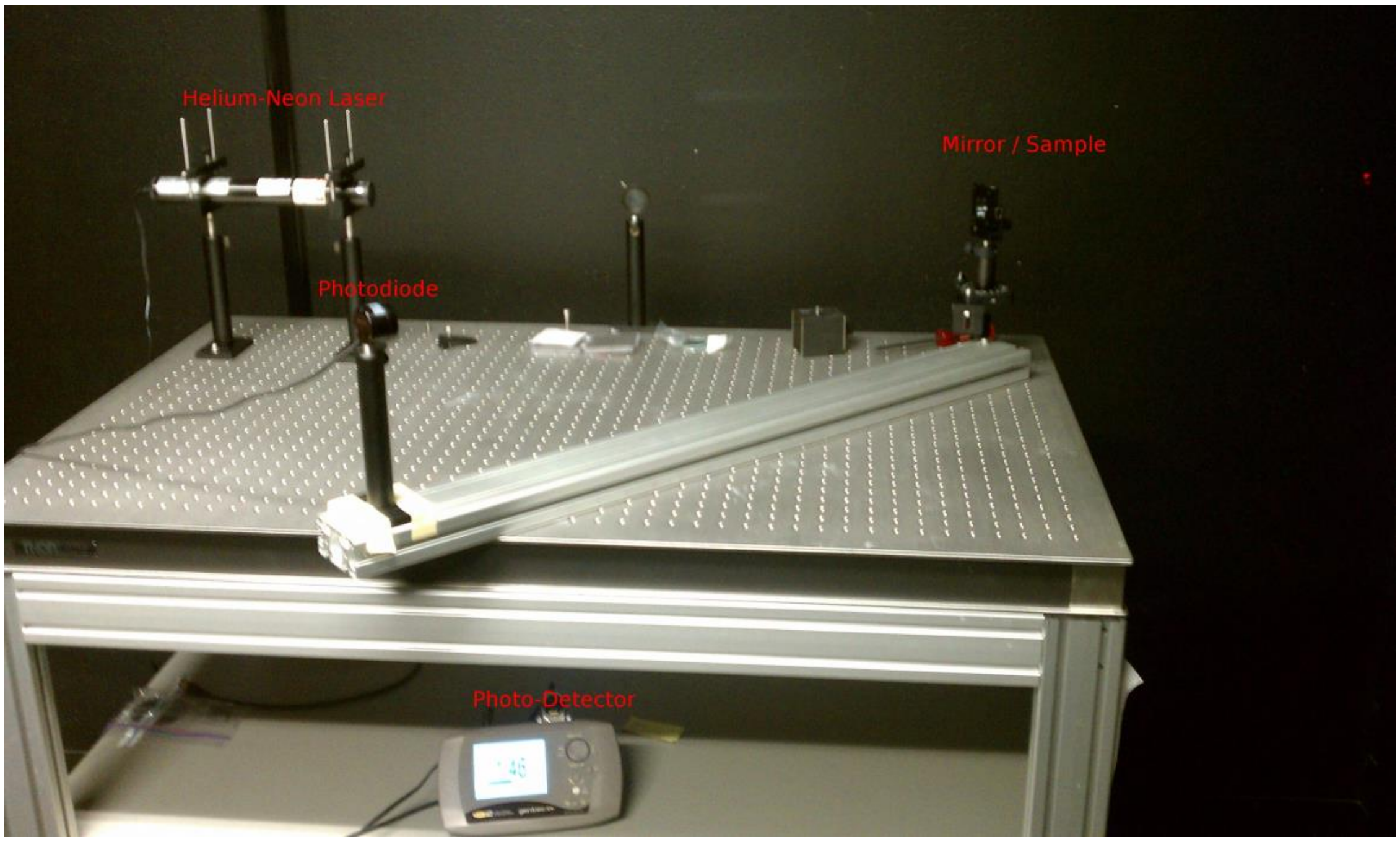

Figure 5. Experimental setup for specular reflectance measurements. The setup consists of a Helium-Neon laser (top-left) which is reflected by the sample (top-right) and sent towards the photodiode detector (middle-left) where an intensity measurement is taken using the Gentec photo-detector readout (bottom-middle). The arm on which the photodiode is mounted is articulated to obtain measurements at each of the three angles.

The procedure used to obtain these measurements is as follows. The laser is aligned using a mirror to ensure that it is incident on the center of both the angled sample and the photodiode. A measurement of the initial intensity of the laser is obtained. Each sample is measured at each of the three angles. The alignment of the laser is reconfirmed once the three measurements are taken. From these measurements, the specular reflectance is calculated by taking the average of the power incident on the photodiode measured at each of the three angles and dividing by the initial intensity of the laser. Finally, the ratio of the specular to total reflectance of each sample is computed to determine each sample's fraction of specular reflectance. The total reflectance is simply the sum of the diffuse and specular reflectance; the diffuse reflectance fraction of each sample is simply 1 minus the specular reflectance fraction.

### 4.2 Specular reflectance measurements

Figure 6 and Figure 7 give the ratio of specular to total reflectance for many of the samples with total reflectance measurements presented the previous section.

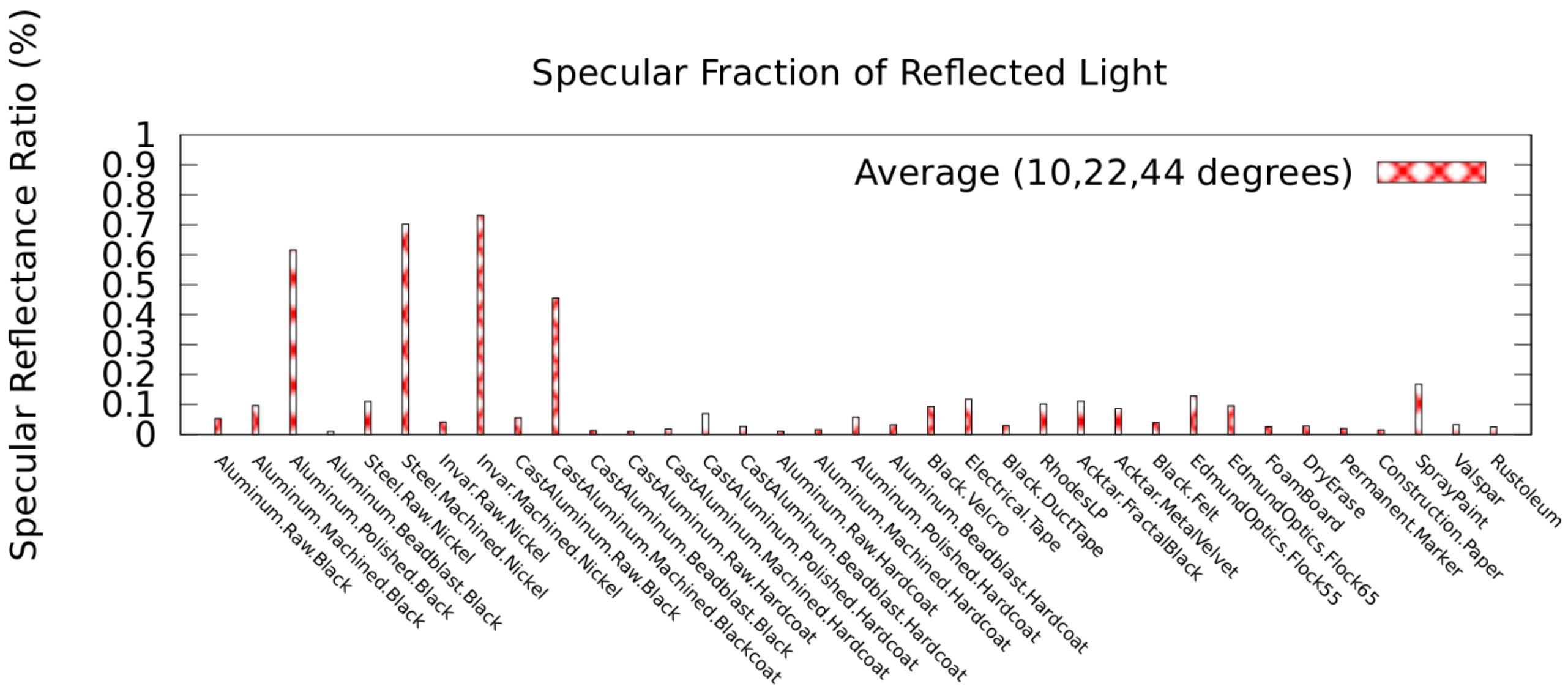

Figure 6. Specular reflectance of a subset of the samples.

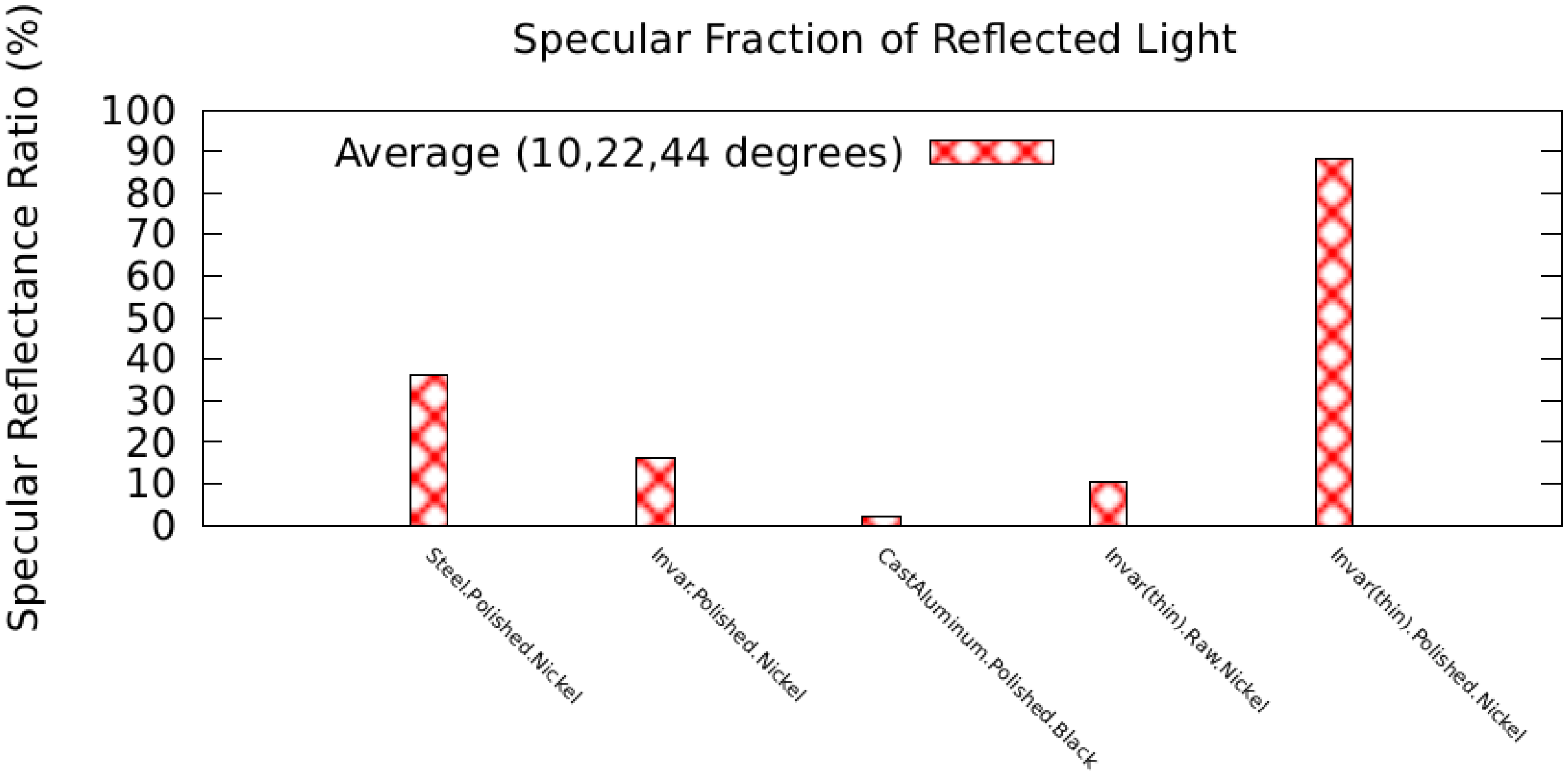

Figure 7. Specular reflectance of a subset of the samples.

## 5. DISCUSSION

The total reflectance measurements presented in Section 2 show several interesting results. Black anodization is frequently used to treat all surfaces inside astronomical instruments; the measurements presented here show, however, that black anodization is quite reflective at red and near-IR wavelengths ($\lambda > 700$ nm), and in fact non-dyed (“hardcoat”) anodization of aluminum may be a better treatment for instruments that cover a wide wavelength range, particularly if the surfaces are not beadblasted prior to coating.

We find that nearly all of the modern commercially available blackout materials that we tested do in fact have very low total reflectance across the entire wavelength range studied here. Before these modern materials were commonly available, however, simpler materials were often used as blackening agents within astronomical instruments. Much to the authors' dismay and consternation, we find that black construction paper, which has often been used in the past in an attempt to minimize stray light within infrared instruments, is in fact not an appropriate blackout material for optical systems designed for $\lambda > 600$ nm. Similarly, black permanent marker is shown to be an unsuitable blackening agent for all but very blue wavelength regions.

The specular reflectance measurements presented in Section 3 also produce some noteworthy results. Of the various metal surface treatments studied here (raw, machined, polished, and beadblasted), we show that beadblasting is (perhaps not surprisingly) the preferred surface treatment to mitigate specular reflection of light by black anodized aluminum surfaces. Beadblasting also produces a surface with the lowest total reflectivity for black anodized aluminum. Interestingly, while the particular surface finish of black anodized 6061 and cast aluminum (and also of nickel plated steel and Invar) has a significant effect on the specularity of the light reflected by the surface, the same is not true for hardcoat non-dyed anodized surfaces which do not show a strong variation in specularity with surface finish. In fact, for wavelengths $\lambda > 700$ nm, for all but beadblasted surfaces, the total reflectance of non-dyed anodization is significantly smaller than that of the black anodization ubiquitously used in astronomical instruments. This result suggests that it is preferable to use non-dyed rather than black anodized aluminum coatings in astronomical instruments in the future.

It is also interesting to note that while the final surface finish of both Invar and steel surfaces shows a similar variation in specularity, the total reflectance of these surfaces (when treated with an electroless nickel coating) is very low at all wavelengths, so the surface finish may not be particularly important.

## 6. CONCLUSIONS

We have presented measurements of the amount of total and specular reflectance of various materials that have been or may be used to minimize stray and scattered light within optical and near-infrared astronomical instruments. We have shown some expected, and also some surprising, properties of these black materials. In particular, the appearance of blackness or shininess to the human eye is not necessarily a measure of the appropriateness of a material for minimizing scattered light along an optical path. If reflection of stray light by surfaces within an instrument is a significant concern, appropriate blackening of metal and non-metal surfaces within the instrument should be carefully considered.

## ACKNOWLEDGEMENTS

Texas A&M University thanks Charles R. '62 and Judith G. Munnerlyn, George P. '40 and Cynthia Woods Mitchell, and their families for support of astronomical instrumentation activities in the Department of Physics and Astronomy.